\documentstyle[aps,epsf]{revtex}
\input amssym.def
\def\xref#1{(\ref{#1})}
\newcommand{\calD}{{\cal{D}}}

\newcommand{\calH}{{\cal{H}}}

\newcommand{\calT}{{\cal{T}}}

\begin{document}
%\date{\today}
\title{Chaos in the Kepler
System}
\author{
C. Chicone\thanks{ Department of  Mathematics,
University of Missouri, Columbia, MO 65211.
Supported by the NSF grant DMS-9531811 and the 
University of Missouri Research Board.} ,\quad
B. Mashhoon\thanks{ Department of  Physics and Astronomy,
University of Missouri, Columbia, MO 65211.
}  \quad
and D. G. Retzloff\thanks{ Department of Chemical 
Engineering,
University of Missouri, Columbia, MO 65211.}
}

\maketitle

\begin{abstract}
The long-term dynamical evolution of a Keplerian binary
orbit due to the emission and absorption of gravitational
radiation is investigated.
This work extends our previous results on
transient chaos in the planar case to the
{\it three dimensional} Kepler system.
Specifically, we consider the nonlinear evolution of the
relative
orbit due to gravitational radiation damping as well as
external gravitational radiation that is {\it obliquely}
incident on the initial orbital plane.
The variation of orbital inclination, 
especially during resonance capture, turns out to be very 
sensitive to the initial conditions.
Moreover, we discuss the novel phenomenon of chaotic transition.
\end{abstract}

\pacs{0430,0540,9510C,9780}

\section{Introduction}
In a recent investigation of
the radiative perturbations of
the {\em planar} Kepler problem
\cite{1}, we have found
evidence for {\em transient
chaos}. Specifically, we have
considered in our previous work~\cite{1,2} an isolated binary
system consisting of masses
$m_1$ and $m_2$ whose internal
structures have been ignored,
i.e. they have been treated
essentially as Newtonian point
masses with positions ${\bf x}_1$ 
and ${\bf x}_2$ in an
inertial frame. The relative
motion is damped since orbital
energy and angular momentum
leave the system due to the
emission of gravitational
radiation. Moreover, the binary
system is perturbed by an
external normally incident
plane gravitational wave. All
radiative perturbations of the
binary system have been treated
in the quadrupole approximation
in our work
\cite{1}; therefore, it follows
that the center-of-mass motion
is unaffected by the radiative
effects. The well-known
connection between the Kepler
system and the harmonic
oscillator implies that the
system under consideration is
analogous to a certain damped
oscillator with external
periodic forcing. In fact, it
is expected that replacing the
emission and absorption of
gravitational radiation by
other similar damping and
forcing mechanisms,
respectively, might lead to
similar phenomena. For
instance, the influence of the
external radiation could be
replaced by the tidal
perturbations of a distant
third mass.

The transient chaos in the  dynamical evolution of the
relative orbit under radiative
perturbations appears to be associated with
the phenomenon of capture into
resonance as explained in detail in our present work. Moreover,
we point out in this paper an interesting aspect of transient chaos,
namely, the phenomenon of chaotic transition.  The
slow evolution of the
binary orbit while locked in
resonance can be explained
using the second order averaged
dynamics \cite{2} that is described here for the three
dimensional Kepler system.

It is important to emphasize the
idealized  nature of the model under
investigation here \cite{1,2,3,4}. We
consider the simplest Keplerian model
in which the effects of emission and
absorption of gravitational radiation
are taken into account. The radiative
damping is of particular
interest in view of the timing
observations of the
Hulse-Taylor relativistic binary pulsar
PSR~B1913+16: the data can be
interpreted in terms of energy loss
due to the emission of gravitational
radiation in agreement with general
relativity \cite{5,6}. For recent
discussions of the measurable
relativistic effects in binary
systems, we refer to the
investigations of Kopeikin and his
collaborators \cite{7}.

It is estimated that about half of all
stars are members of binary or
multiple systems; therefore, it is
possible that the general phenomena we
have encountered in our theoretical
investigations have actually occurred
in nature. In particular, our results
\cite{1,2,3,4} should be relevant for
the behavior of relativistic binary
pulsars; indeed, a number of such
systems have been discovered since the
existence of the original binary
pulsar PSR~B1913+16 was first
recognized~\cite{5,8}. This
possibility provides the impetus to
analyze the Kepler system under more
general conditions, especially in
connection with the nature of its
chaotic behavior. To this end, we
consider in this paper the radiative
perturbations of the {\em three
dimensional} Kepler system. As work
continues on relativistic binary
systems---following the Taylor-Hulse
work on the original binary
pulsar---and more are discovered and
studied, it may be that the
theoretical results presented in our
work could be useful in the
elucidation of observed phenomena in
such interesting astronomical systems.

\section{Dynamical Equations}

The simplest equation of relative
motion for the Kepler system including
radiative perturbations is given by
\begin{equation}\label{eq1}
\frac{d^2x^i}{dt^2}+k
\frac{x^i}{|{\bf x}|^3}+{\cal{R}}^i=
-\epsilon {\cal{K}}_{ij}(t)x^j,
\end{equation}
where ${\bf x}:={\bf x}_1-{\bf x}_2$
is the vector describing relative
motion, $k:=G_0 M$, $M:=m_1+m_2$, the
quantity $-{\bf{\cal{R}}}$ is
the relative acceleration
caused by gravitational
radiation reaction, and
$\epsilon \cal{K}$ is the
tidal matrix of the external
wave evaluated at the center of
mass of the binary system that
we take to be the origin of spatial
coordinates in the background
inertial frame. Thus, in this 
frame the motion of $m_1$ and $m_2$ can be  described by
${\bf x}_1(t)=(m_2/M) {\bf x}(t)$
and ${\bf x}_2(t)=-(m_1/M) {\bf x}(t)$, respectively.
Various properties of the system~
\xref{eq1} have been discussed in
detail in our recent investigations of
the planar Kepler  problem
\cite{1,2,3,4}; therefore, we
concentrate in this paper on the long
term deviations of the system~\xref{eq1} from planar motion.

In the quadrupole
approximation, the standard
expression for radiation
damping may be reduced---for
the Kepler system under
consideration---to
\begin{equation}
{\cal{R}}=
\frac{4G_0^2m_1m_2}{5c^5|{\bf
x}|^3}[\psi {\bf v}
   -\chi (\hat{\bf x}\cdot {\bf
v})\hat{\bf x}] \: \:,\label{eq2}
\end{equation}
where ${\bf v}=d{\bf x}/dt$ is
the relative velocity,
$\hat{\bf x}={\bf x}/|{\bf x}|$, and
\begin{equation}\label{eq3}
\psi=12v^2-30(\hat{\bf x}\cdot {\bf
v})^2-\frac{4k}{|{\bf x}|} \: \:,
  \quad \chi =36v^2-50(\hat{\bf
x}\cdot {\bf v})^2+
    \frac{4k}{3|{\bf x}|} \:.
\end{equation}
Moreover, the tidal matrix for a plane
monochromatic wave of frequency
$\Omega$ propagating in the
$(x^1,x^3)$-plane along a direction
that makes an angle $\Theta$, $0\le
\Theta\le \pi$, with respect to the
$x^3$-axis is given by (cf. appendix A)
\begin{eqnarray}
{\cal{K}}_{11}&=&\alpha
\Omega^2\cos^2 \Theta \cos \Omega
t,\nonumber\\ 
{\cal{K}}_{12}&=&\beta
\Omega^2\cos \Theta \cos (\Omega
t+\phi_0),\nonumber\\
{\cal{K}}_{13}&=&-\alpha
\Omega^2\cos \Theta \sin \Theta
\cos \Omega t,\nonumber\\
{\cal{K}}_{22}&=&-\alpha
\Omega^2\cos \Omega t,\nonumber\\
{\cal{K}}_{23}&=&-\beta
\Omega^2\sin \Theta \cos (\Omega
t+\phi_0),\label{eq4}
\end{eqnarray}
and the other components of the tidal
matrix are determined by the fact that
$({\cal{K}}_{ij})$ is symmetric and
traceless. Here $\alpha $ and $\beta$
are amplitudes of the two independent
linear polarization states of the
incident radiation and
$\phi_0$ is the constant phase
difference between them. The
background inertial frame is fixed by
the form of the tidal matrix
\xref{eq4} and the fact that the
initial unperturbed orbit is assumed
to be in the
$(x^1,x^2)$-plane. The strength of the
external perturbation is given by
$\epsilon$, $0<\epsilon <<1$, so that
$\alpha$ and $\beta$ are of the order
of unity; at present, it is expected
that $\epsilon \sim 10^{-20}$ though
gravitational waves have not yet been
detected in the laboratory.

It must be emphasized that the
simple form of the external
perturbation assumed here is meant to
represent the dominant
component of a wave packet
composed mostly of wavelengths
much larger than the size of
the Keplerian system. In the
absence of the external
perturbation, an orbit in the
$(x^1,x^2)$-plane would remain
confined to this plane under
radiative damping. If the
external wave is turned on at a
certain instant, it is expected
that the equation of motion
\xref{eq1} would represent the
steady-state situation after
transients have died away.

It is interesting to subject
the equation of motion to
certain scale transformations
that would render it
dimensionless. To this end, let
${\bf x}=R_0\,\tilde{\bf x}$ and
$t=T_0\,\tilde{t}$ such that
$R_0$ and $T_0$ are constants
and a tilde denotes a
dimensionless quantity. In this
way the equation of motion
keeps its form except that $k$
must be replaced by $\tilde{k}$,
\begin{equation}\label{eq5}
\tilde{k}=k\frac{T_0^2}{R_0^3} \: \: ,
\end{equation}
$\Omega =\tilde{\Omega}/T_0$
and the strength of the
gravitational radiation
reaction is given by the
dimensionless quantity $\delta$,
\begin{equation}\label{eq6}
\delta =\frac{4G_0^2m_1m_2}{5c^5
T_0R_0} \: \:.
\end{equation}
In the
rest of this paper, we fix the
relationship between $R_0$ and
$T_0$ by $kT_0^2=R_0^3$, so that
$\tilde{k}=1$. Thus if $R_0$ is
taken to be the semimajor axis
of the initial Keplerian orbit,
then its period is given by
$2\pi T_0$.
We adopt this convention regarding the initial orbit throughout this work.
It follows that
$\delta <(20\sqrt{2})^{-1}$ in
all physically significant
cases; in fact, $\delta \sim
10^{-15}$ for the binary pulsar
PSR~B$1913+16$ \cite{5,6}. Henceforth
we deal with dimensionless
equations of motion in which
all tildes are dropped.

It is appropriate to introduce
cylindrical coordinates
$(\rho,\theta, z)$ such that
$x^1=\rho\cos \theta $, $x^2=\rho\sin
\theta $, and $x^3=z$. The
equations of motion can then be
expressed in the form

\begin{eqnarray}
\frac{d\rho}{dt}&=&p_{\rho},
\quad
\frac{d\theta}{dt}=
\frac{p_\theta}{\rho^2},\quad
\frac{dz}{dt}=p_z,\nonumber\\
\frac{dp_{\rho}}{dt}&=&-
\frac{\rho}{(\rho^2+z^2)^{3/2}}+
\frac{p_\theta^2}{\rho^3}-
{\cal{R}}^{\rho}\nonumber\\
&&\quad-\epsilon
\rho\left[\frac{1}{2}
({\cal{K}}_{11}+
{\cal{K}}_{22})+
\frac{1}{2}({\cal{K}}_{11}-
{\cal{K}}_{22})\cos 2\theta
+{\cal{K}}_{12}\sin 2\theta
\right]\nonumber\\
&&\quad-\epsilon
z({\cal{K}}_{13}\cos \theta
+{\cal{K}}_{23}\sin \theta
),\nonumber\\
\frac{dp_\theta}{dt}&&=-
{\cal{R}}^\theta -\epsilon
\rho^2\left[{\cal{K}}_{12}\cos
2\theta
-\frac{1}{2}({\cal{K}}_{11}-
{\cal{K}}_{22})\sin 2\theta
\right]\nonumber\\
&&\quad -\epsilon
\rho z(-{\cal{K}}_{13}\sin \theta
+{\cal{K}}_{23}\cos \theta
),\nonumber\\
\frac{dp_z}{dt}&=&-
\frac{z}{(\rho^2+z^2)^{3/2}}-
{\cal{R}}^z-\epsilon
\rho({\cal{K}}_{13}\cos \theta
+{\cal{K}}_{23}\sin \theta
)-\epsilon
z{\cal{K}}_{33},
\label{eq7}
\end{eqnarray}
where we use equation \xref{eq4} for
the external perturbation and
\begin{equation}\label{neq8}
{\cal{K}}_{33}=-({\cal{K}}_{11}+{\cal{K}}_{22})=\alpha \Omega^2\sin^2\Theta\cos \Omega t.
\end{equation}
In equation \xref{eq7}, the radiation
reaction terms are given by

\begin{eqnarray}
{\cal{R}}^{\rho}&=&
\frac{\delta }{(\rho^2+z^2)^{3/2}}
  \left[\psi p_{\rho}-\chi
\frac{\rho(\rho\, p_{\rho}+zp_z)}{\rho^2+z^2}
\right],\nonumber\\
{\cal{R}}^\theta
&=&\frac{\delta
\psi p_\theta}{(\rho^2+z^2)^{3/2}},
\nonumber\\
{\cal{R}}^z&=&
\frac{\delta}{(\rho^2+z^2)^{3/2}}
\left[\psi p_z-\chi 
\frac{z(\rho\, p_{\rho}+zp_z)}{\rho^2+z^2}
\right],\label{eq8}
\end{eqnarray}
where $\psi$ and $\chi$ are given by
equation \xref{eq3} and can be
expressed as
\begin{eqnarray}
\psi&=&12\left(p^2_{\rho}+\frac{p_\theta
^2}{\rho^2}+p_z^2\right)-30
\frac{(\rho\, p_{\rho}+zp_z)^2}{\rho^2+z^2}-
\frac{4}{(\rho^2+z^2)^{1/2}},
\nonumber\\
\chi&=&36\left(p^2_{\rho}+\frac{p_\theta^2}{\rho^2}+p_z^2\right)
-50 \frac{(\rho\, p_{\rho}+zp_z)^2}{\rho^2+z^2}+
\frac{4}{3(\rho^2+z^2)^{1/2}}.
\label{eq9}
\end{eqnarray}
It now
remains to integrate these
equations with appropriate
boundary conditions.

\section{Numerical Experiments}

The equations of motion \xref{eq7}
have been so formulated that for
$\Theta = 0$, we recover the results
of our recent work \cite{1,2}.  We
wish to investigate how our previous
results vary with $\Theta$ as the
direction of incidence of the external
radiation deviates from the normal. It
is clear that our previous theoretical
results,  such as regarding the
persistence of periodic orbits,  would
still hold for sufficiently small
$\Theta $, 
$0<\Theta \ll 1$; however, the analysis
for arbitrary
$\Theta$ would be cumbersome and we
therefore resort  to numerical
experiments. To express our numerical
results in a form that could be easily
compared with previous results
\cite{1,2}, we note that the energy
$E$, total angular momentum $G$ and
the $z$-component of the angular
momentum $H$ for the osculating
ellipse in our system can be expressed
in terms of cylindrical coordinates as
\begin{eqnarray}\label{DynEqCyl}
E &=& \frac{1}{2}
\left(p_{\rho}^2+\frac{p_{\theta}^2}
{\rho^2}+p_z^2\right) -
\frac{1}{(\rho^2+z^2)^{1/2}}, \\ 
G &=&
\left[(\rho\, p_z-z
p_{\rho})^2+\left(1+\frac{z^2}
{\rho^2}\right)
p_{\theta}^2\right]^{1/2}, \\ 
H &=& p_{\theta},
\end{eqnarray}
respectively.  The corresponding
Delaunay elements are the action
variables $(L, G, H)$, where
$L=(-2E)^{-1/2}$. The osculating
ellipse is depicted in figure~\ref{oef}.

Some typical results of our numerical
experiments are presented in
figures~\ref{capfig1}--\ref{capfig5}.
Figure~\ref{capfig1} explores the
phenomenon associated with {\em capture into resonance};
here the osculating orbit has
eccentricity
$e$, $G = L(1-e^2)^{1/2}$, and $H = G
\cos{i}$, where $i$ denotes the
orbital inclination. The angle of
incidence is $\Theta = \pi/10$. 
Figures~\ref{capfig2}--\ref{capfig5}
explore the phenomena associated with {\em transient chaos}
for $\Theta = \pi/6$.

To arrive at these representative results, we begin our numerical
work with a resonant orbit given at some initial instant of time
by 
\[(p_\rho, p_\theta,p_z,\rho,\theta,z)=(e_0,1,0,1,0,0),\]
which corresponds to an osculating ellipse in the
$(x^1,x^2)$-plane with eccentricity $e_0$, unit orbital
angular momentum about the $z$-axis, semimajor axis $(1-e_0^2)^{-1}$
and Keplerian frequency $(1-e_0^2)^{3/2}$. 
In figure~\ref{oef}, for instance, this initial orbit would have
$h=0$, $g=-\pi/2$ and $\hat{v}=\pi/2$. We then assume
an external wave that is a coherent superposition of the two
independent linear polarization states with equal amplitudes
$(\alpha=\beta)$ and zero phase shift between them $(\phi_0=0)$.
The wave is incident at an angle $\Theta$ that we vary in our experiments.
We choose the wave frequency $\Omega=m\omega$ to correspond
to the $(m:1)$ resonance. Integrating forward and 
backward from this initial system, we explore the region
around the resonance configuration. Starting from a state before
resonance capture, we then integrate forward
until the integration routine gives up due to the rapid collapse of the
binary system. To compare our results with the planar case $\Theta=0$,
we choose the two planar configurations with initial $e_0=1/2$ that 
we studied in a previous paper~\cite{1}. In the first system, i.e. 
figure~1 of~\cite{1}, we used
$\epsilon=10^{-4}$, $\delta=10^{-7}$, $\alpha=\beta=2$,
$\phi_0=0$ and $\Omega=L_0^{-3}$ with $L_0=(1-e_0^2)^{-1/2}$.
This $(1:1)$ resonance is studied for $\Theta=\pi/10$ in 
figure~\ref{capfig1} of the present paper. In the second system, i.e.
figure~2 of~\cite{1}, we used $\epsilon=10^{-3}$, 
$\delta=10^{-8}$, $\alpha=\beta=2$,
$\phi_0=0$ and $\Omega=2 L_0^{-3}$ with $L_0= (1-e_0^2)^{-1/2}$. 
This $(2:1)$ resonance is studied for $\Theta=\pi/6$ in 
figures~\ref{capfig2}--\ref{capfig5} of the present paper; in these figures, every
tenth iterate of the $2\pi/\Omega$ stroboscopic Poincar\'e map is
plotted. This second case exhibits transient chaos and to be reasonably
certain of the various features of this chaotic regime, we have used
two different standard routines for stiff integration of~\xref{eq7}
using double precision arithmetic. The results of the integration
of this configuration are
depicted in figures~\ref{capfig2} and~\ref{capfig3}. These
are to be compared with figures~\ref{capfig4} and~\ref{capfig5}, 
respectively, which depict the results of integration of 
essentially the same system albeit with a different starting point.
It is important
to note that the orbital inclination is particularly sensitive to transient
chaos as is clear from the behavior of $\cos i$ in 
figures~\ref{capfig3} and~\ref{capfig5}.
Moreover, as the binary system collapses, we expect that eventually
the binary orbit would tend to a circle ($e\to 0$). 
In figures~\ref{capfig2}--\ref{capfig5} the endpoint of
integration exhibits a sharp drop in eccentricity as would be theoretically
expected. Note that this behaviour of eccentricity coincides with the sharp
bend in $L(t)$ near $L=1$ corresponding to chaotic transition as
described in section~\ref{sec:5}.

In the following sections, we present a
theoretical explanation of some of the
main features of the behavior of the
system over a long period of time. For
this purpose, essential use is made of
the dynamical equations in terms of
Delaunay elements (cf. appendix B).

\section{Averaging}

The long-term behavior of the dynamical
system under investigation is best
revealed after we average the system
over the Keplerian frequency $\omega =
L^{-3}$. In fact, the Delaunay
equations (cf. appendix B)
 show clearly that the three
dimensional Kepler system has only one
intrinsic frequency, namely,
$\omega$.  Let us write the equations
of motion in the form
\begin{eqnarray}\label{EQM1}
\frac{dL}{dt} &=& -\epsilon \frac{\partial {\cal H}^*}
{\partial \ell} + \epsilon \Delta f_L, \nonumber \\
\frac{dG}{dt} &=& -\epsilon \frac{\partial {\cal H}^*}
{\partial g} + \epsilon \Delta f_G, \nonumber \\
\frac{dH}{dt} &=& -\epsilon \frac{\partial {\cal H}^*}
{\partial h} + \epsilon \Delta f_H, \nonumber \\
\frac{d\ell}{dt} &=& \frac{1}{L^3}+\epsilon
\frac{\partial {\cal H}^*}{\partial L} +
\epsilon \Delta f_{\ell}, \nonumber \\
\frac{dg}{dt} &=& \epsilon \frac{\partial
{\cal H}^*}{\partial G} + \epsilon \Delta f_g, \nonumber \\
\frac{dh}{dt} &=& \epsilon \frac{\partial {\cal H}^*}
{\partial H} + \epsilon \Delta f_h, \nonumber \\
\frac{d{\tau}}{dt} &=& \Omega,
\end{eqnarray}
where ${\cal H}^*$ is the Hamiltonian associated
with the external perturbation, $\Delta=\delta/\epsilon$ and $f_D$,
$D\in \{ L, G, H, \ell, g, h\}$, denotes the
gravitational radiation damping.  Here $\tau =
\Omega t$ is a new angular variable.

In the absence of the external perturbation,
we have shown \cite{1} that the system
maintains its orbital plane---i.e., the
inclination of the orbit remains fixed---but loses
energy and angular momentum
so that the orbit decays monotonically
and eventually collapses.  During this
process the orbit tends to a circle, i.e.
the eccentricity $e$ decreases.  With
 the external radiation present, as in
\xref{eq1}, we have a dynamical system
with two frequencies $\omega$ and $\Omega$
and the possibility of resonance needs
to be taken into account.  This would
occur if relatively prime integers $m$
 and $n$ exist such that $m \omega = n
\Omega$; the $(m:n)$ resonance manifold
is then given by the set
\[\{(L, G, H, \ell, g, h, \tau)\: :
\: m\frac{1}{L^3} = n\Omega \}.\]
Let us first assume that the dynamical
system \xref{EQM1} is off resonance;
then, averaging \xref{EQM1} over a period
 of the external radiation results in a
 system that is simply damped. That is,
the terms involving ${\cal H}^*$ in
\xref{EQM1} simply drop out; therefore,
 the system tends to a circular
orbit as it decays on the average
 while maintaining the inclination
of the orbital plane (cf. appendix B).  This explains
figure~\ref{capfig1} before and after resonance
capture; that is, in these regimes $e$
decreases monotonically on the average
while $\cos{i}$ is essentially constant.
If the resonance condition is satisfied,
then the system either passes through
the resonance without being captured
or is captured into the resonance.
In the former case, the system behaves on the average
in the same way that it would off resonance.
In the latter case, however, the resonance
condition fixes the semimajor axis
of the osculating ellipse {\it on
the average}.  Thus the external wave
deposits energy into the system so as
to balance radiation damping on the
average and hence $L$ remains fixed
on the average while the exchange of
angular momentum takes place.
The resulting change in $G$ and $H$ affects the
shape as well as the configuration of the orbit; 
in fact, the eccentricity
and the inclination of the orbit vary
while its semimajor axis oscillates about
the value fixed by the resonance.  This is clearly
indicated in figure~\ref{capfig1}, where the
inclination and the eccentricity of the
orbit both decrease as a result of passage
through resonance.  The rate of damping
depends on the shape of the orbit and
this fact accounts for the difference
in the slope of $L(t)$ before and after
capture into resonance. During resonance capture, 
we let $L=L_{0}+\epsilon ^{1/2}{\cal D}$ and
$\ell=\phi +n\Omega t/m$, where $L_0^{-3}=n\Omega /m$ 
and ${\cal D}$ and $\phi$ are the new canonical
variables associated with resonance.
The
behavior of the dynamical system while
trapped in resonance can be studied using
the method of partial averaging (cf.
appendix C). It turns out that for the
radiative perturbations of the
three dimensional Kepler system under
consideration here resonance is possible
only for $n=1$. Numerical experiments then
reveal that the $(1:1)$ resonance has a
simple structure as in figure 2. More
complicated structures are expected for
higher order resonances as described in
our previous planar work \cite{2}. In the present paper, we explore
a $(2:1)$ resonance in figures 3-6. A
general discussion of the
$(m:1)$ resonance for the Kepler system is
beyond the scope of this work.  

To explain the behavior of the system
while trapped in resonance in a
quantitative manner, we need to develop
the first and second order partially averaged
dynamics for \xref{EQM1} following the
method adapted in our recent work on the
planar Kepler system~\cite{2}.  
This is
done in appendix C and the resulting
system of second order partially averaged
Delaunay equations for the $(m:1)$ resonance are
\begin{eqnarray}
\dot{{\cal D}} &=&
-\epsilon ^{1/2}\left[ -m\calT_c\sin m\phi
+m\calT_s\cos m\phi
+\frac{\Delta}{G^7}\left(8
+\frac{73}{3}e^2+\frac{37}{12}e^4
\right)\right]\nonumber\\ 
&&\quad-\epsilon {\cal D}\left[
-m\frac{\partial
\calT_c}{\partial L} \sin m\phi
+m\frac{\partial
\calT_s}{\partial L}\cos  m\phi 
+\frac{\Delta}{3L^3_0G^5}(146+37e^2)\right],
\nonumber\\
\dot G &=& -\epsilon \left[
\frac{\partial \calT_c}{\partial g}\cos
m\phi +\frac{\partial
\calT_s}{\partial g} \sin m\phi +
\frac{\Delta
}{L^3_0G^4}(8+7e^2)\right],\nonumber\\
\dot{H} & = & -\epsilon \left[
\frac{\partial \calT_c}{\partial h} \cos
m\phi +\frac{\partial
\calT_s}{\partial h} \sin m\phi
+\frac{\Delta }{L^3_0G^4}(8+7e^2)\cos
i\right],\nonumber\\
\dot{\phi} &=& -\epsilon ^{1/2}
\frac{3}{L^4_0}{\cal D} +\epsilon \left(
\frac{6}{L^5_0}{\cal D}^2+\frac{\partial
{\cal T}_c}{\partial L}\cos m\phi
+\frac{\partial {\cal T}_s}{\partial
L}\sin m\phi \right),\nonumber\\
\dot{g} & = & \epsilon \left(
\frac{\partial {\cal T}_c}{\partial G}\cos
m\phi +\frac{\partial {\cal T}
_s}{\partial G} \sin m\phi
\right),\nonumber\\
\dot{h} & = & \epsilon \left(
\frac{\partial {\cal T}_c}{\partial H}
\cos m\phi +\frac{\partial{\cal T}
_s}{\partial H}\sin m\phi \right),
\label{eq15}
\end{eqnarray}
where $\calT_c$ and ${\calT}_s$ are
complicated expressions involving $\alpha
,\beta,\Omega ,\phi_0,\Theta$ and the
Delaunay elements given by (C25) in
appendix C. It is important to note that
the first-order averaged equations are
formally the same as in \cite{2}; that is,
to order $\epsilon ^{1/2}$ the orbital
inclination is fixed and we simply have an oscillator in $\calD$ and
a pendulum with constant torque in $\phi$ as before.
To second order in
$\epsilon^{1/2}$, the damping (or
antidamping) term enters the pendulum
equation and this results in an antidamped
(or damped) harmonic oscillator in ${\cal
D}$ as in the case illustrated in figure
2. These features have been explained in
detail in \cite{2}; therefore, it is more
interesting to investigate the new aspects
of resonance capture associated with the
slow oscillation of the plane of the
osculating orbit. In fact, the possibility
of the variation of the inclination to
second order in $\epsilon^{1/2}$ would
result in an additional feature that is
required for a complete explanation  of
figure~\ref{capfig1}. The pendulum in $\phi$ will
couple, in the general case, to an
equation for the variation in the orbital
inclination. To develop this latter
equation, we note that $H=G\cos i$ and
hence 
$d(\cos i)/dt=(\dot{H}G-H\dot{G})/G^2$,
where
$\dot{G}$ and
$\dot{H}$ are given by \xref{EQM1}. We
are interested in the slow variation in
the inclination of the orbit during
resonance capture. The average behavior of
the orbital inclination is obtained if we
substitute for $\dot{G}$ and $\dot{H}$
from the second order partially averaged
system~\xref{eq15} instead. In this way,
we find 
\begin{equation}
\label{eq16} 
\frac{d}{dt}\cos i =-\frac{\epsilon
}{G}\left[ \left(
\frac{\partial {\cal T}_c}{\partial
h}-\frac{\partial {\cal T}_c}{\partial g}
\cos i\right) \cos m\phi +\left(
\frac{\partial {\cal T}_s}{\partial
h}-\frac{\partial {\cal T}_s}{\partial
g}\cos i\right) \sin m\phi
\right],
\end{equation} 
for the slow variation of
the orbital inclination {\em while the
system is captured in resonance.} It is
remarkable that the explicit damping terms
proportional to $\Delta$ in \xref{eq15}
drop out in equation \xref{eq16}; indeed, in
the absence of external waves there would be no
resonance capture and $\cos i$ would be constant. Thus the rate of
variation of $\cos i$ is directly proportional to the presence of the 
external perturbation as in \xref{eq16}. It is difficult to draw exact
conclusions regarding $\cos i$ from \xref{eq15} and \xref{eq16}; however,
we intuitively expect that the orbital
inclination would undergo on average
simple oscillatory movements during resonance capture. These
oscillations are expected to be {\it slow}, 
since $\cos i$ varies in time over
the temporal scale given by $\epsilon t$.
These conclusions are generally consistent
with the results of our numerical
experiments. 

\section{Transient Chaos}\label{sec:5}

Numerical experiments suggest that our
system exhibits chaotic transients near
the ``exit from resonance''. In our
experiments, we considered the
$(2:1)$ resonance similar to the planar case in figure 2 of
\cite{1}. Starting
at the point
$(p_\rho,p_\theta,p_z,\rho,\theta,z)=(0.5, 1, 0, 1, 0, 0)$
which is near the resonance manifold,
backward integration shows that the system
is captured into resonance; then, after
a sufficiently long forward integration,
the orbit leaves the vicinity of the
resonance and continues toward collapse.
However, the length of time required for
the system to exit the resonance is very
sensitive to the integration method and to
changes in the initial conditions. Also,
plots of $L$ versus $t$ show that the
signal appears to pass through a chaotic
region after it finally
exits the resonance. This strongly
suggests that the system is passing
through an unstable chaotic set; that is,
there is transient chaos.

We expect transient chaotic effects to be present
throughout the resonance capture region; however,
in our experiments transient chaos appears most prominently
after exit from resonance. Following the exit from the $(2:1)$
resonance under consideration, the
system collapses relatively slowly until
it undergoes a certain {\it chaotic
transition} and the relative orbit that
emerges after this transition collapses
much more rapidly. In fact, the system
makes a transition from one relative
orbit to a totally different relative
orbit in a rather short period of time
while undergoing  what appears to be transient chaos. For lack
of a better characterization of this
unexpected phenomenon, we refer to it as
``chaotic transition''. As a consequence
of this transformation, a ``bent knee''
appears in figures~\ref{capfig2}--\ref{capfig5} near $L=1$; 
indeed, the
rate at which the system  collapses
suddenly changes at the ``bend''. That is, the
orbital eccentricity increases very
rapidly and this leads to a more rapid
rate of loss of energy due to the emission
of gravitational radiation---since this is
proportional to $(1-e^2)^{-7/2}$ as can be
seen from inspection of equation
(15) of \cite{1}---and hence the ``bend.''
We have verified by two different numerical methods that these chaotic
phenomena are stable features of the $(2:1)$ resonance.

\section{Conclusion}

We consider in this paper a three
dimensional Kepler system that is subject
to radiative perturbations caused by
radiation damping as well as an incident
monochromatic gravitational wave.

A Keplerian binary system constantly loses
energy to gravitational radiation
according to general relativity. There is
a lack of complete reciprocity between the
emission and absorption of gravitational
radiation, however. The absorption of
gravitational radiation by the binary is
not monotonic and the system sometimes
gains energy from the external wave and
sometimes deposits energy into the wave.
The behavior of the system averaged over
the period of the external wave is thus
one of continuous collapse due to
radiation damping except when it is
captured into resonance. Other than at
resonance, the {\em average} behavior of
the system is that the orbital plane
remains fixed while the system shrinks as
it tends to a circle $(a\to 0,e\to 0)$. At
resonance, the system on average steadily
{\em gains} energy from the external wave
in order to balance the steady loss of
energy to radiation. The resonance
condition $m\omega =n\Omega$ is necessary
but not sufficient for the occurrence of
this delicate balance at the $(m:n)$
resonance. In general, transient chaotic
behavior is expected near a resonance. As
the system evolves, it passes through and
is affected by a {\em dense} set of $(m:n)$ resonances. 
When it is indeed captured
into a resonance, we find that $n=1$.

We have not proven the existence of
transient chaos in the Kepler system when
radiative perturbations are taken into
account. Instead, we have presented
further numerical evidence in support of
this hypothesis in this paper. Two
different codes for the numerical
integration of equation~\xref{eq7} have
been employed and they have yielded
qualitatively the same results. Our
conclusions regarding transient chaos and
chaotic transition are based on these
numerical results for the three dimensional Kepler system.

The parameter space of the problem under
consideration is very large. No significant attempt has
been made at exploring this parameter
space; for instance, we cannot exclude the
possibility of existence of a strange
attractor in our system. Instead, we have
concentrated our attention on a $(1:1)$
resonance and a $(2:1)$ resonance in the three dimensional case that are
familiar from our previous planar work~\cite{1,2}. The $(1:1)$ resonance has a
simple structure and its general features
can be explained by the second order
averaged dynamics. That is, the semimajor
axis undergoes antidamped oscillations
about its resonance value while the
orbital inclination slowly oscillates
until the system leaves the resonance; on
the other hand, angular momentum is
transferred to the binary orbit during
resonance such that the orbit after the
resonance is substantially different in
eccentricity from the orbit before it gets
captured into resonance and hence so is
the rate of collapse of the binary due to
the emission of gravitational radiation.

The $(2:1)$ resonance has a rich
structure
and much of our numerical analysis has
been concerned with its elucidation. 
It appears that this structure is associated with transient chaos.
Once the system leaves the resonance, the
general trend towards collapse is
accompanied by complex structure that we
characterize as transient chaotic
behavior. This is followed by a peculiar
chaotic transition  which may be another manifestation of
transient chaos. The result is that the system
collapses extremely rapidly as a result of
going through this transition. Integrating
backward in time from the resonance, the
system grows in size. Since the radiation
reaction force decreases with size at
least as $r^{-5}$ while the tidal force of
the external wave grows as $r$, the end
result is that the radiation reaction can
become effectively negligible. This is the
regime of Arnold diffusion in our problem
and can lead to gravitational ionization
as described in detail in our previous
work \cite{3,4}.

It is possible that for the general
$(m:1)$ resonance other interesting
phenomena may occur that are beyond the
scope of this investigation.

\appendix

\renewcommand{\theequation}{A\arabic{equation}}

\section{Absorption of Gravitational Waves}

Consider a plane monochromatic
gravitational wave with wave vector ${\bf
K}$ and frequency $\Omega =c|{\bf K}|$.
The spacetime metric is given by $g_{\mu
\nu}=\eta_{\mu\nu}+\epsilon h_{\mu\nu}$,
where we impose the gauge condition
$h_{0\mu}=0$. Here $\eta_{\mu\nu}$ is the Minkowski metric with
signature $+2$; moreover, it is important to note that our
gauge has the characteristic property that the
worldline of a static test particle is a geodesic and this is consistent with
the binary system as a whole remaining at rest in this external radiation
field. 
The gravitational wave
amplitude is then characterized by a
symmetric and traceless $h_{ij}$ with
$\partial _jh_{ij}=0$ and $\Box h_{ij}=0$.
Let
\begin{equation}
\label{eqA1}
h_{ij}={\text Re}\,[\hat{h}_{ij}\exp (-i\Omega
t+i{\bf K}\cdot {\bf x})],
\end{equation}
where $\hat{h}_{ij}K^j=0$. 
We are interested in the gravitational influence of this wave on a Keplerian
binary system. If the wavelength of the radiation is much larger than 
the dimension of the system, then the interaction of the wave with the
system is predominantly tidal in character.
Then the tidal
matrix is given by
\begin{equation}\label{eqA2} 
{\cal
K}_{ij}=-\frac{1}{2}
\frac{\partial^2h_{ij}}{\partial
t^2}(t,{\bf 0 }),
\end{equation} 
since the center of mass of the binary
system is fixed at the origin of the
spatial coordinates. We choose the
coordinate system such that the wave is
incident in the $(x,z)$-plane; hence,
$\hat{\bf K}=(\sin \Theta ,0,\cos \Theta)$
is the unit propagation vector of the
wave. Let us then choose $(\hat{\bf
K},\hat{\bf N},\hat{{\bf y}})$ to be an
orthonormal triad with $\hat{\bf N}=(\cos \Theta , 0,-\sin \Theta )$. 
Expanding the symmetric and
traceless $\hat{h}_{ij}$ in terms of this
basis triad and taking the transversality
of the wave into account
$(\hat{h}_{ij}\hat{K}^j=0)$, we find
\begin{equation}\label{eqA3} 
\hat{h}_{ij}=2\alpha(\hat{N}_i\hat{N}_j-\hat{y}_i\hat{y}_j)
  +2\beta\exp (-i\phi_0)(\hat{N}_i\hat{y}_j+\hat{N}_j\hat{y}_i), 
\end{equation} 
where
$2\alpha$ and $2\beta$ are the real
amplitudes of the two independent linear
polarization states and
$\phi_0$ is the constant phase difference
between them. Here the zero of time is so
chosen that $\alpha$ is real. It is now
possible to recover equation~\xref{eq4}
from \xref{eqA1}-\xref{eqA3}.

The absorption of tidal gravitational energy
from such a wave is not unidirectional in
general; that is, a system can gain or
lose energy as a result of its tidal
interaction with an incident gravitational
wave. On the other hand, we have found
that during resonance capture the system
always {\em gains} energy.

The nonlinear evolution of the binary
system has been emphasized in our recent
work when the external perturbation can,
in effect, be considered to be a plane
monochromatic gravitational wave~\cite{1,2,3,4}. 
However, we have not
considered an arbitrary incident wave
packet or a stochastic background of
gravitational waves. These cases have been
treated in the absence of radiation
damping for the {\em linear} evolution of
the binary system \cite{9}. It turns out
that for a binary immersed in a random,
isotropic and unpolarized linear
gravitational wave background
characterized solely by its energy
spectral density, the osculating elements
of the relative orbit undergo random walks
\cite{9}. The extension of these results
to the long term nonlinear evolution of
the system remains a task for the future.
In this connection, let us mention that nonlinear stochastic differential
equations have been the subject of numerous investigations 
(see, for example,~\cite{10,11}).

\section{Equations of Motion in Delaunay's Variables}

\renewcommand{\theequation}{B\arabic{equation}}

\vspace*{.25in}

The equations of motion in terms of
Delaunay's action-angle variables are
of basic importance for the dynamical
considerations in this paper.  The
standard treatments usually involve
 Hamiltonian systems; therefore, we
present here a direct derivation of
these equations since our system is
dissipative.

The equations of motion for the perturbed
Kepler problem \xref{eq1} can be written
in vector form as
\begin{equation}\label{vkepler}
\frac{d^2 {\bf x}}{dt^2}
 + \frac{{\bf x}}{r^3} = {\bf F},
\end{equation}
where $r = |{\bf x}|$ and ${\bf F}$
is given by
\begin{eqnarray}\label{ForcingF}
F^i & = & -\epsilon{\cal K}_{ij} x^j - 
\frac{\delta}{r^3}\left[\left(12v^2-30
\dot{r}^2 -\frac{4}{r}\right)v^i \right.
\nonumber\\ 
& & \left.
-\frac{\dot{r}}{r}\left(36v^2-50\dot{r}^2
+\frac{4}{3r}\right)x^i\right].
\end{eqnarray}
At any instant of time $t$, ${\bf x}$ and
${\bf v}$ specify the instantaneous orbital
plane that contains the osculating ellipse
with its focus always at the origin of
coordinates.
 Let us choose the coordinate system such
that the osculating ellipse at time $t$
has the form given in figure~\ref{oef}.

To obtain the equations of motion in
Delaunay elements, we employ an orthonormal
frame field adapted to the instantaneous
osculating ellipse, write the component
equations of motion using this basis,
and perform  a transformation
from the inertial Cartesian coordinates 
to those adapted to the instantaneous
osculating ellipse.
We define the
frame field as follows.  Let
$\hat{\bf r}$ be a unit radial vector,
${\bf x} = r \hat{\bf r}$, and
$\hat{\bf n}$ be a unit vector normal to
the plane of the instantaneous osculating ellipse.  
To complete this frame field,
we include $\hat{\bf s} = \hat{\bf n}
\times\hat{\bf r}$ which is a unit vector
in the orbital plane normal to $\hat{\bf
r}$.
Then ${\bf F}$ can be written as
\begin{equation}
\label{ForceFrame}
{\bf F} = F_r \hat{\bf r}+F_s \hat{\bf s}+
F_n \hat{\bf n}
\end{equation}
using its radial, sideways and normal components.
In what follows, we use the definition of
 the Delaunay elements $(L, G, H, \ell, g, h)$ given by
\begin{eqnarray}\label{Deldefn}
L &=& a^{1/2}, \nonumber \\
G &=& L (1-e^2)^{1/2}, \nonumber \\
H &=& G \cos{\:i}, \nonumber \\
\ell &=& \hat{u} - e \sin{\:\hat{u}},
\nonumber \\
g &=& \mbox{\rm argument of the periastron},
\nonumber \\
h &=& \mbox{\rm longitude of the ascending
node},
\end{eqnarray}
where $a$ is the semimajor axis of the
osculating ellipse, $e$ is its eccentricity,
$\hat{u}$ is the eccentric anomaly, and
$\ell$ is the mean anomaly.  Only positive 
square roots are considered throughout this paper. The equation
for the radial position $r$ in terms of the
true anomaly $\hat{v}$ is given by
\begin{equation}\label{rveqn}
r=\frac{a(1-e^2)}{1+e \cos{\hat{v}}}.
\end{equation}
The orbital energy \xref{DynEqCyl} is given by
\begin{equation}\label{DynEq1}
E = \frac{1}{2}v^2 - \frac{1}{r} = -\frac{1}{2a},
\end{equation}
where \[v^2 = \dot{r}^2 + \frac{a(1-e^2)}{r^2}\]
using the fact that $G = |{\bf
x}\times{\bf v}|$. It follows from these
relations and \xref{rveqn} that
\begin{equation}\label{rdot}
\dot{r} =\frac{e\sin{\hat{v}}}{G}.
\end{equation}
It is clear from \xref{vkepler} that
\begin{equation}\label{energy1}
\frac{dE}{dt} = {\bf F}\cdot{\bf v}.
\end{equation}
It then follows from \xref{rdot}, \xref{energy1},
and the definition of $L$ that we have
\begin{equation}\label{Lequation}
\frac{dL}{dt} = L^3 \: {\bf F}\cdot{\bf v} =
\frac{a}{(1-e^2)^{1/2}}\left[F_r e \sin{\hat{v}}
+F_s\frac{a(1-e^2)}{r}\right].
\end{equation}
To obtain the equations of motion for
the remaining Delaunay elements, we use the relations
\[
{\bf x}\times{\bf v} = G {\bf{\hat n}},\qquad
\frac{d}{dt}({\bf x}\times{\bf v}) = {\bf x}
\times{\bf F}
\]
to get
\begin{equation}\label{Geqn}
\frac{dG}{dt} = rF_s,
\end{equation}
and
\begin{equation}\label{SecondGeqn}
G\frac{d\hat{\bf n}}{dt} = -rF_n\:\hat{\bf s}.
\end{equation}
Now consider a coordinate system $(x',y',z')$
oriented to the instantaneous osculating ellipse
such that the perturbed motion is given by
$x' = r \cos{(\hat{v}+g)}$, $y' = r \sin{(\hat{v}+g)}$,
and $z'=0$.  The transformation from the coordinate system
 $(x, y, z)$ to $(x',y',z')$ consists of a rotation
by an angle $h$ about the $z$-axis followed by a
rotation by the inclination angle $i$  about the
line of the ascending node, which is the $x'$-axis.
This transformation can be written as
\begin{equation}\label{transformation}
\left( \begin{array}{c}
x \\ y \\ z
\end{array} \right)  =
\left( \begin{array}{ccc}
\cos{h} & -\sin{h}\cos{i} & \sin{h}\sin{i} \\
\sin{h} & \cos{h}\cos{i} & -\cos{h}\sin{i} \\
0 & \sin{i} & \cos{i}
\end{array}\right)
\left( \begin{array}{c}
x' \\ y' \\ z'
\end{array} \right).
\end{equation}
It follows that the relative position is given by 
\begin{eqnarray}\label{CoordinateTrans}
x = r \sin{\vartheta}\cos{\varphi} & = &
r[\cos{h}\cos{(\hat{v}+g)}-\sin{h}\cos{i}
\sin{(\hat{v}+g)}], \nonumber \\ y = r
\sin{\vartheta}\sin{\varphi} & = &
r[\sin{h}\cos{(\hat{v}+g)}+\cos{h}\cos{i}
\sin{(\hat{v}+g)}], \nonumber \\ 
z = r \cos{\vartheta} & = & r
\sin{i}
\sin{(\hat{v}+g)}.
\end{eqnarray}
In the $(x', y', z')$ coordinate system,
$\hat{\bf n} = (0, 0, 1)$.
Thus in the inertial frame,
\[
\hat{\bf n} = (\sin{h}\sin{i}, -\cos{h}\sin{i}, \cos{i}),
\]
and
\begin{eqnarray}\label{n1eqn}
\hat{\bf s} &=& \hat{\bf n}\times\hat{\bf r}\nonumber\\
&=& (-\cos{h}\sin{(\hat{v}+g)}-\sin{h}\cos{i}\cos{(\hat{v}+g)},
\nonumber \\ 
&&\hspace{.5in}\cos{i}\cos{h}\cos{(\hat{v}+g)}-
\sin{h}\sin{(\hat{v}+g)}, \sin{i}\cos{(\hat{v}+g)}).
\end{eqnarray}
Substituting \xref{n1eqn} into \xref{SecondGeqn}
results in 
\begin{eqnarray}\label{Delaunay34}
\frac{di}{dt} & = & \frac{rF_n}{G}
\cos{(\hat{v}+g)}, \nonumber \\
\: \frac{dh}{dt} \sin{i} & = &
\frac{rF_n}{G}\sin{(\hat{v}+g)}.
\end{eqnarray}
{}From the definition of $H$ in
\xref{Deldefn} and equations~\xref{Geqn} and~\xref{Delaunay34},
it follows that
\begin{equation}\label{Heqn}
\frac{dH}{dt} = r[F_s\cos{i}
-F_n\sin{i}\cos{(\hat{v}+g)}].
\end{equation}
The dynamical equation for the mean
anomaly is obtained by exactly the
same method as used in~\cite{1} (cf.
appendix B in~\cite{1}).  The result for the three
dimensional case is obtained by replacing
$F_{\theta}$ in~\cite{1}  by $F_s$; hence,
\begin{equation}\label{elleqn}
\frac{d\ell}{dt} = \omega + \frac{r}{e(a)^{1/2}}
[F_r(-2e+\cos{\hat{v}}
+e\cos^2{\hat{v}})-F_s(2+e\cos{\hat{v}})
\sin{\hat{v}}].
\end{equation}
To obtain the equation for the dynamics
of the argument of the periastron,
consider the relative velocity vector in the form
\[{\bf v} = \dot{r} \hat{\bf r} + \frac{G}{r}
\hat{\bf s}.\]
Project onto the $z$-axis to obtain
\begin{equation}\label{zdoteqn}
\dot{z} = \dot{r}\sin{i}\sin{(\hat{v}+g)}
+\frac{G}{r}\sin{i}\cos{(\hat{v}+g)}.
\end{equation}
Differentiating the $z$-component of
\xref{CoordinateTrans} with respect to
time and equating the result to \xref{zdoteqn} gives
\begin{equation}\label{vgeqn}
\frac{d(\hat{v}+g)}{dt} = \frac{G}{r^2}
-\frac{rF_n}{G}\frac{\cos{i}}{\sin{i}}
\sin{(\hat{v}+g)}.
\end{equation}
Now, by taking the time derivative of the
logarithm of~\xref{rveqn}, we obtain
\begin{equation}\label{v1eqn}
\frac{d \hat{v}}{dt} = \frac{G}{r^2}
- \frac{G}{e} \left[-F_r\cos{\hat{v}}
+F_s\left(1+\frac{r}{a(1-e^2)}\right)\sin{\hat{v}}
\right].
\end{equation}
Subtracting \xref{v1eqn} from
\xref{vgeqn} gives the dynamical
equation for the argument of periastron as
\begin{eqnarray}\label{geqn}
\frac{dg}{dt} &=& -\frac{rF_n}{G}\frac{\cos{i}}{\sin{i}}
\sin{(\hat{v}+g)} \nonumber \\
&&+\frac{G}{e} \left[-F_r\cos{\hat{v}}
+F_s\left(1
+\frac{r}{a(1-e^2)}\right)\sin{\hat{v}}
\right].
\end{eqnarray}
To summarize, the dynamical equations in Delaunay elements are
\begin{eqnarray}\label{Tdynamical eqn}
\frac{dL}{dt} & = &  \frac{a}{(1-e^2)^{1/2}}
\left[F_r e \sin{\hat{v}}
+F_s\frac{a(1-e^2)}{r}\right], \nonumber \\
\frac{dG}{dt} &=& rF_s, \nonumber \\
\frac{dH}{dt} &=& r[F_s\cos{i}
-F_n\sin{i}\cos{(\hat{v}+g)}], \nonumber \\
\frac{d\ell}{dt} &=& \omega
+ \frac{r}{e(a)^{1/2}}[F_r(-2e+\cos{\hat{v}}
+e\cos^2{\hat{v}})
-F_s(2+e\cos{\hat{v}})\sin{\hat{v}}], \nonumber \\
\frac{dg}{dt} &=& -\frac{rF_n}{G}
\frac{\cos{i}}{\sin{i}}\sin{(\hat{v}+g)} \nonumber \\
&&+\frac{(a(1-e^2))^{1/2}}{e} \left[-F_r\cos{\hat{v}}
+F_s\left(1+
\frac{r}{a(1-e^2)}\right)\sin{\hat{v}}
\right], \nonumber \\
\frac{dh}{dt} & = &  \frac{rF_n}{G}
\frac{\sin{(\hat{v}+g)}}{\sin{i}}.
\end{eqnarray}
It now remains to express $F_r$, $F_s$ and $F_n$
in terms of Delaunay elements.

The equations of motion in Delaunay's variables
can be expressed in the form given in \xref{EQM1}.  
To this end,
let
\begin{equation}\label{hstardefn}
{\cal H}^* = \frac{1}{2} {\cal K}_{ij} x^i x^j
\end{equation}
be the Hamiltonian associated with the
external perturbation and $\delta = \epsilon \Delta$.
Then~\xref{ForcingF} can be written as
\begin{equation}\label{eqB24} 
{\bf F}=-\epsilon \nabla {\cal H}^*+\epsilon
\Delta {\bf f},
\end{equation}
where 
\begin{equation}\label{eqB25} 
{\bf f}=\frac{\dot{r}}{r^3}\left(
24v^2-20\dot{r}^2+\frac{16}{3r}\right)\hat{\bf r}
-\frac{G}{r^4}\left( 12v^2-30\dot{r}^2-\frac{4}{r}\right)
\hat{\bf s}.
\end{equation}
Thus $f_r=\dot{r} (\chi -\psi )/r^3$, $f_s=-G\psi /r^4$ and
$f_n=0$. It is important to note that the functions $\psi$ and
$\chi$ can be expressed solely in terms of Delaunay variables
in the plane of the orbit $(L,G,\ell,g)$ just as in our previous
work \cite{1,2}. It then follows from the inspection of
equations
\xref{eqB24} and \xref{eqB25} that when radiation damping acts
alone, the orbital plane remains fixed in space while the orbit
tends to a circle as it shrinks $(e\to 0$ and $a\to 0$). Using
these results in
\xref{Tdynamical eqn}, one can show that equation~\xref{EQM1}
is recovered with appropriate definitions of the damping
functions
$f_D$ and the replacement of $\Omega t$ by $\tau$.

\section{Partial Averaging}

\renewcommand{\theequation}{C\arabic{equation}}

\vspace*{.25in}

This appendix is devoted to the determination of the {\em
resonant} behavior of the three dimensional Kepler
system when averaged over the binary
period. Averaging over this ``fast'' motion should reveal the
``slow'' oscillatory behavior of the system while in resonance
and its exit out of the resonance.

The perturbed Kepler system continuously emits gravitational
waves and hence loses energy and angular momentum to
gravitational radiation damping. Hence the semimajor axis of
the osculating ellipse must steadily decrease except when the
system is trapped in a resonance. At resonance, a delicate
balance exists between the external gravitational perturbation
and the radiation damping since the energy (and hence the
semimajor axis) of the osculating ellipse would be constant on
the average. That is, during resonance capture the energy
deposited into the orbit by the external wave would equal on
average the energy lost via the emission of gravitational
radiation. In general, the system slowly drifts out of the
resonance when this subtle balance is significantly upset.

Let $L=L_0$ when the resonance condition $m\omega=n\Omega $ is
exactly satisfied; then, we assume
\begin{eqnarray}\label{eqC1} 
L&=&L_0+\epsilon ^{1/2}{\cal D},\\
\ell&=&\frac{1}{L^3_0}t+\phi,\label{eqC2}
\end{eqnarray}
during resonance capture. As before \cite{2}, we recast the
equations of motion in Delaunay variables
\xref{EQM1} in terms of the new variables $({\cal D},\phi )$. To
this end, 
\begin{equation}\label{eqC3} 
\frac{1}{L^3}=\frac{1}{L^3_0}\left[ 1-3\frac{{\cal
D}}{L_0}\epsilon^{1/2} +6\frac{{\cal D}^2}{L^2_0}\epsilon
+{\cal O} (\epsilon^{3/2})\right],
\end{equation} 
and so we
obtain from \xref{EQM1} upon Taylor expansion
\begin{eqnarray}\label{eqC4} 
\dot{\cal D}&=&-\epsilon ^{1/2} F_{11}-\epsilon {\cal
D}F_{12},\nonumber\\
\dot{G} &=& -\epsilon F_{22},\nonumber\\
\dot{H}&=& -\epsilon F_{32},\nonumber\\
\dot{\phi}& =&-\epsilon^{1/2}\frac{3{\cal D}}{L^4_0}
    +\epsilon\Big(\frac{6\calD^2}{L_0^5}+F_{42}\Big),\nonumber\\
\dot{g}&=& \epsilon F_{52},\nonumber\\
\dot{h} & = & \epsilon F_{62},
\end{eqnarray}
where terms of order higher than $\epsilon$ have been neglected. Here $(F_{ij})$ are defined by
\begin{eqnarray}\label{eqC5} 
F_{11} & = & \frac{\partial {\calH}^*}{\partial \ell}-\Delta
f_L,\nonumber\\ 
F_{12} & =& \frac{\partial ^2{\calH}^*}{\partial L\partial \ell} 
   -\Delta \frac{\partial f_L}{\partial L},\nonumber\\ 
F_{22} &=& \frac{\partial {\calH}^*}{\partial g} 
   -\Delta f_G,\nonumber\\ 
F_{32} & = &
\frac{\partial {\cal H}^*}{\partial h}-\Delta
f_H,
\end{eqnarray} 
while
\begin{eqnarray}\label{eqC6} 
F_{42} & = & \frac{\partial {\calH}^*}{\partial L} +\Delta
f_\ell,\nonumber\\ 
F_{52} & = & \frac{\partial {\calH}^*}{\partial
G}+\Delta f_g,\nonumber\\ 
F_{62} & = & \frac{\partial {\calH}^*}{\partial H}+\Delta f_h.
\end{eqnarray} 
Note that the first index in $F_{ij}$ refers to the equation in
which it appears while the second index refers to its order in
powers of the perturbation parameter $\epsilon^{1/2}$.
Moreover, all the terms on the right sides of \xref{eqC5} and
\xref{eqC6} are to be evaluated at $(L_0,G,H,\phi +n\Omega
t/m,g,h)$ so that
$(F_{ij})$ become functions of $(G,H,\phi +n\Omega t/m,g,h,t)$,
since ${\cal H}^*$ is explicitly dependent upon time. Let us further
note that at resonance the appropriate perturbation parameter
turns out to be
$\epsilon^{1/2}$, since only with this choice would the
conjugate variables ${\cal D}$ and $\phi$ both vary
predominantly on the same slow temporal scale given by
$\epsilon^{1/2}t$.

The system \xref{eqC4} is periodic in $t$ with period $2\pi
m/\Omega$. To average this system over its period, we introduce
an averaging transformation that has the function of rendering
the transformed system
\xref{eqC4} in a form that is averaged to first order in
$\epsilon^{1/2}$. The resulting system will then be replaced by
the second order averaged system, where the terms of order
$\epsilon$ are simply averaged over the $2\pi m/\Omega$ period.
The averaging theorem ensures that the solution of the second
order averaged system is sufficiently close to the solution of
\xref{eqC4} over a timescale of order $\epsilon^{-1/2}$
\cite{2}.

Let us first define the average of $F_{ij}$ to be 
\begin{equation}\label{eqC7} 
\langle F_{ij}\rangle =\frac{\Omega }{2\pi m}\int_0^{2\pi
m/\Omega }F_{ij}(G, H, \frac{n\Omega }{m} t+\phi , g, h, t)\, dt.
\end{equation} 
Since the averaging transformation involves only first order
quantities, the only term of interest in
\xref{eqC4} would then be $F_{11}$ and we let $\lambda (G, H,
\phi ,g,h, t)=F_{11}-\langle F_{11}\rangle $ be its deviation
from the average. Next, we define $\Lambda (G,H, \phi ,g, h,
t)$ to be the antiderivative of $\lambda$ with respect to $t$
such that $\langle \Lambda \rangle=0$. The averaging
transformation is then given by
\begin{equation}\label{eqC8} 
{\cal D}=\hat{\cal D}-\epsilon ^{1/2}\Lambda (\hat{G}, \hat{H},
\hat{\phi} ,\hat{g},\hat{h}, t),
\end{equation}
$G=\hat{G}$, $H=\hat{H}$, $\phi =\hat{\phi}$, $g=\hat{g}$ and
$h=\hat{h}$. Hence the averaging transformation is {\em on average}
equivalent to the identity transformation. Upon this
transformation,
\xref{eqC4} takes the form
\begin{eqnarray}\label{eqC9} 
\dot{\hat{\cal D}} & = & -\epsilon ^{1/2} \langle F_{11}\rangle
-\epsilon
\hat{\cal D} \left( \frac{3}{L^4_0}\frac{\partial
\Lambda}{\partial \hat{\phi}}+F_{12}\right),\nonumber\\
\dot{\hat{G}}&=&-\epsilon F_{22},\nonumber\\
\dot{\hat{H}}&=&-\epsilon F_{32},\nonumber\\
\dot{\hat{\phi}} & = & -\epsilon ^{1/2} \left(
\frac{3}{L^4_0}\hat{\cal D}\right) +\epsilon \left(
\frac{6}{L^5_0}\hat{\cal D}^2+\frac{3}{L^4_0}\Lambda
+F_{42}\right),\nonumber\\
\dot{\hat{g}} & = & \epsilon F_{52},\nonumber\\
\dot{\hat{h}} & = & \epsilon F_{62},
\end{eqnarray}
where terms of order $\epsilon ^{3/2}$ and higher have been
neglected. We note that \xref{eqC9} is in averaged form to first
order, but not to second order. Therefore, we now average
\xref{eqC9} to obtain the second order partially averaged system
that we wish to study. To simplify the notation, we write our
main result in the form
\begin{eqnarray}\label{eqC10} 
\dot{\cal D} &=& -\epsilon ^{1/2} \langle F_{11}\rangle
-\epsilon {\cal D}
\langle F_{12}\rangle,\nonumber\\
\dot{G} & =& -\epsilon \langle F_{22}\rangle, \nonumber\\
\dot{H} & =& -\epsilon \langle F_{32}\rangle, \nonumber \\
\dot{\phi} & = & -\epsilon^{1/2}\left(\frac{3}{L^4_0}{\cal D}\right)+\epsilon \left( \frac{6}{L^5_0}{\cal
D}^2+\langle F_{42}\rangle\right),\nonumber\\
\dot{g} & =& \epsilon \langle F_{52}\rangle ,\nonumber\\
\dot{h} & =& \epsilon \langle F_{62}\rangle ,
\end{eqnarray}
since $\langle\Lambda \rangle =0$ and $\langle \partial \Lambda
/\partial \phi \rangle =0$. We remark that
\xref{eqC10} is simply the averaged form of our original system
\xref{eqC4}; however, this will not be true in general but
happens to be the case here. It now remains to compute the
averages $\langle F_{ij}\rangle$ defined by \xref{eqC7}.

Let us first focus attention on the radiation reaction terms.
It follows from the discussion in the last paragraph of
appendix B that the effect of
this frictional force is to maintain the orbital plane while
the relative orbit loses energy and shrinks $(a\to 0,e\to 0)$.
Therefore, the result of averaging is essentially the same as
in our previous work \cite{1,2} and is given by 
\begin{eqnarray}\label{eqC11} 
\langle f_L\rangle &=& -\frac{1}{G^7}\left( 8 +
\frac{73}{3}e^2+\frac{37}{12}e^4\right),\\
\langle f_G\rangle & =&-\frac{1}{L^3G^4}(8+7e^2),\label{eqC12}\\
\langle f_H\rangle &=& \langle f_G\rangle \cos
i,\label{eqc13}
\end{eqnarray} 
and $\langle f_\ell\rangle =\langle
f_g\rangle =\langle f_h\rangle =0$. Moreover, $\langle \partial
f_L/\partial L\rangle =\partial \langle f_L\rangle /\partial L$
and hence
\begin{equation}\label{eqC14} 
\langle \frac{\partial f_L}{\partial L}\rangle
=-\frac{1}{3L^3G^5}(146+37e^2).
\end{equation}
All these quantities are to be evaluated at $L=L_0$.

We consider next the terms involving the external Hamiltonian
$\epsilon {\cal H}^*$, which must first be expressed in terms
of the Delaunay elements. To this end, let
\begin{eqnarray}\label{eqC15}
S(h)&=&\sin h,\nonumber\\
U(G,H,h) & =& \cos \Theta \cos i\; S+ \sin \Theta \sin i,\nonumber\\
V(G,H,h)&=& \cos h \cos i,\nonumber\\
W(h) & = & \cos \Theta \cos h,
\end{eqnarray}
and define $P_\sigma$ and $\tilde{P}_\sigma $, $\sigma =0,\pm$, such
that
\begin{eqnarray}
\label{eqC16} 
P_0&=& \frac{1}{2}\left(
-S^2+U^2-V^2+W^2\right),\nonumber\\ 
P_+ & = & \frac{1}{2}
\left( -S^2-U^2+V^2+W^2\right),\nonumber\\ 
P_- & =&
-SV-UW,\nonumber\\
\tilde{P}_0&=& SW-UV,\nonumber\\
\tilde{P}_+ & = & SW+UV,\nonumber\\
\tilde{P}_-&=& -SU+VW.
\end{eqnarray}
Using these quantities, we can form ${\cal P}_\sigma (G,H,h,t)$,
\begin{equation}\label{eqC17} 
{\cal P}_\sigma
=\frac{1}{2}\alpha \Omega ^2\cos \Omega t\;P_\sigma
+\frac{1}{2}\beta
\Omega^2\cos (\Omega t+\phi_0)\tilde{P}_\sigma .\end{equation}
It is now possible to express ${\cal H}^*$ in terms of the
Delaunay variables as follows:
\begin{equation}\label{eqC18} 
{\cal H}^*=\sum_{\sigma =0,\pm}{\cal P}_\sigma {\cal Q}^\sigma ,
\end{equation} 
where 
\begin{eqnarray}\label{eqC19} 
{\cal Q}^0 (L,G,\ell,g)&=&r^2=a^2(1+\frac{3}{2}e^2)
 -4a^2\sum_{\nu=1}^\infty \frac{1}{\nu^2}J_\nu(\nu e)\cos\nu\ell, \\
{\cal Q}^+ (L,G,\ell,g)&=&r^2\cos
(2\hat{v}+2g)=\frac{5}{2}a^2e^2\cos 2g+a^2\sum^\infty_{\nu
=1}(A_\nu \cos 2g\cos \nu \ell -B_\nu \sin 2 g\sin \nu \ell
),\nonumber\\ 
{\cal Q}^-(L,G,\ell ,g)&=& r^2\sin (2\hat{v}+2g)=
\frac{5}{2}a^2e^2\sin 2g+a^2\sum^\infty_{\nu =1}(A_\nu
\sin 2g\cos \nu \ell +B_\nu \cos 2g\sin \nu \ell ).\nonumber\\
&&{}
\end{eqnarray}
Here $A_\nu$ and $B_\nu$ are functions of the eccentricity
$e=(L^2-G^2)^{1/2}/L$ and can be expressed in terms of the Bessel
function $J_\nu (x)$ as
\begin{eqnarray}\label{eqC20} 
A_\nu & = & \frac{4}{\nu^2 e^2}[2\nu e(1-e^2)J'_\nu (\nu
e)-(2-e^2)J_\nu (\nu e)],\nonumber\\
B_\nu & = & -\frac{8}{\nu^2e^2}(1-e^2)^{1/2}[eJ'_\nu (\nu
e)-\nu (1-e^2)J_\nu (\nu e)],
\end{eqnarray}
where
$J'_\nu (x)=dJ_\nu (x)/dx$. 

We are interested in the average value of ${\cal H}^*$;
therefore, it is convenient to express
${\cal P}_\sigma$ as
\begin{equation} 
\label{eqC21} {\cal P}_\sigma ={\cal C}_\sigma \cos \Omega
t+{\cal S}_\sigma \sin \Omega t,
\end{equation}
where ${\cal C}_\sigma $ and ${\cal S}_\sigma $ can be
explicitly determined using \xref{eqC17}. Moreover, it proves
convenient to express the Fourier series expansions of ${\cal
Q}^\sigma$ collectively as
\begin{equation}\label{eqC22} 
{\cal Q}^\sigma =a^\sigma _0+\sum^\infty_{\nu =1} (a^\sigma_\nu
\cos \nu \ell +b_\nu^\sigma \sin \nu \ell).
\end{equation}
The average of ${\cal H}^*$,
\begin{equation}\label{eqC23} 
\langle {\cal H}^*\rangle =\frac{\Omega}{2\pi m}\int_0^{2\pi
m/\Omega}{\cal H}^* (L,G,H,\frac{n\Omega}{m}t+\phi
,g,h,t)\,dt,
\end{equation} 
is given by
\begin{equation}\label{eqC24} 
\langle {\cal H}^*\rangle ={\cal T} _c(L,G,H,g,h)\cos m\phi
+{\cal T}_s(L,G,H,g,h)\sin m\phi ,
\end{equation}
where
\begin{eqnarray}\label{eqC25} 
{\cal T}_c&=& \frac{1}{2}\sum_\sigma (a_m^\sigma {\cal
C}_\sigma +b_m^\sigma {\cal S}_\sigma ),\nonumber\\
{\cal T}_s&=& \frac{1}{2}\sum_\sigma (-a_m^\sigma {\cal
S}_\sigma +b_m^\sigma {\cal C}_\sigma),
\end{eqnarray} 
for $n=1$. If
$n\neq 1$, then $\langle{\cal H}^*\rangle=0$; hence, we define
${\cal T}_c={\cal T}_s=0$ in this case.

It is now possible to express the quantities in \xref{eqC10}
that depend on the external perturbation in terms of ${\cal
T}_c$ and ${\cal T}_s$. For instance, it is straightforward to
use the formulas
\xref{eqC18}, \xref{eqC22} and \xref{eqC24} to prove that
$$\left\langle\frac{\partial {\cal H}^*}{\partial \ell}\right\rangle =\frac{\partial }{\partial
\phi}\left\langle {\cal H}^*\right\rangle.$$
Moreover, the other partial derivatives of ${\cal H}^*$ with
respect to the other Delaunay variables simply commute with the
operation of averaging. For instance, $\langle \partial {\cal
H}^*/\partial L\rangle =\partial \langle {\cal H}^*\rangle
/\partial L$, which is finally evaluated at $L=L_0$. The second
order partially averaged equations would thus have a form
similar to that for the planar case except for additional
equations for the slow variation of the extra Delaunay
variables $H$ and $h$. More explicitly, the partially averaged
equations for the three dimensional Kepler system are given by
equations~(19) of~\cite{2} (with $L_*\to L_0$, $\varphi
\to \phi$, $T_c\to {\cal T}_c$ and $T_s\to {\cal T}_s$)
together with
\begin{eqnarray*} 
\dot{H} & = & -\epsilon \left[ \frac{\partial {\cal
T}_c}{\partial h} \cos m\phi +\frac{\partial {\cal
T}_s}{\partial h} \sin m\phi +\frac{\Delta
}{L^3_0G^4}(8+7e^2)\cos i\right],\\
\dot{h} & = & \epsilon \left( \frac{\partial {\cal
T}_c}{\partial H} \cos m\phi +\frac{\partial {\cal
T}_s}{\partial H} \sin m\phi \right).
\end{eqnarray*} 
The full
set of partially averaged Delaunay equations is given by
\xref{eq15}.
%\begin{references}

%Fig1
\begin{figure}[h]
\caption{\label{oef} 
Schematic plot of the osculating ellipse
in the three dimensional Kepler problem.
The instantaneous position vector is
${\bf x}(t) = (x, y, z)$.  The unit vector
${\hat{\bf n}}$ is normal to the instantaneous
orbital plane and points in the same direction
as the orbital angular momentum.}
\end{figure}
%Fig2
\begin{figure}[h]
\caption[]{
The plots are for system~(\ref{eq7}) with parameter values
$\epsilon=10^{-4}$, $\delta/\epsilon=10^{-3}$,
$\alpha=2$, $\beta=2$, $\phi_0=0$ and $\Theta=\pi/10$.
Here, $L_0=(4/3)^{1/2}\approx 1.1547$ corresponds to $(1:1)$ resonance with
$\Omega=1/L_0^3$.
The top panel shows $L=a^{1/2}$ versus time $t$ for the initial conditions
$(p_\rho,p_\theta, p_z, \rho,\theta, z)$
equal to
$($0.17554040552969, 0.57754077894430,
         $-0.050059905721582,$
         2.7059982365422,
        $ -6.7506347480153$,
         0.14058165333520$)$.
The middle
panel shows $\cos i$ versus time and the bottom panel shows $e$ versus $t$.
\label{capfig1}}
\end{figure}
%Fig3
\begin{figure}[h]
\caption[]{
The plots are for system~(\ref{eq7}) with the
initial values of $(p_\rho,p_\theta, p_z, \rho,\theta, z)$
equal to\\ $(-$0.6579651108853,
        0.74664767379829,
        $-$0.12269473267506,
        1.4254331835434,
        $-$10.1974793053511,
        $-0.052493593111567)$ 
and parameter values
$\epsilon=10^{-3}$, $\delta/\epsilon=10^{-5}$,
$\alpha=2$, $\beta=2$, $\phi_0=0$, $\Theta=\pi/6$, and 
$\Omega=2/L_0^3$ with $L_0=(4/3)^{1/2}$
so that the initial value is near the $(2:1)$ resonance.
The top panel depicts $L$ versus time, 
the middle panel depicts $G$ versus time,
and the bottom panel depicts the eccentricity $e$ versus time.
Every 10th iterate of the $2\pi/\Omega$ stroboscopic Poincar\'e map is plotted.
\label{capfig2}}
\end{figure}
%Fig4
\begin{figure}[h]
\caption[]{
The top panel depicts $\cos i$ versus time, the middle panel depicts  $G$ versus $L$
and the bottom panel depicts $H$ versus $L$ with the same data as
in figure~\ref{capfig2}.
\label{capfig3}}
\end{figure}
%Fig5
\begin{figure}[h]
\caption[]{
The plots are for system~(\ref{eq7}) with parameters as in figure~\ref{capfig2}
but with initial values of $(p_\rho,p_\theta, p_z, \rho,\theta, z)$
equal to
$(-$0.013456419026, 0.510820155326, 0.161909540107,
2.35369782619, $-644207.714939$, $-2.17211570796)$. 
                                              %$\Omega=1.2990381056766579702$.
The top panel depicts $L$ versus time, the middle panel depicts $G$ versus time,
and the bottom panel depicts the eccentricity $e$ versus time.
Every 10th iterate of the $2\pi/\Omega$ stroboscopic Poincar\'e map is plotted.
\label{capfig4}}
\end{figure}
%Fig6
\begin{figure}[h]
\caption[]{
The top panel depicts $\cos i$ versus time, the middle panel depicts  $G$ versus $L$
and the bottom panel depicts $H$ versus $L$ with the same data as
in figure~\ref{capfig4}.
\label{capfig5}}
\end{figure}
\addtocounter{figure}{-6}
\begin{figure}[h]
\epsfxsize=300pt
\makebox{\epsffile{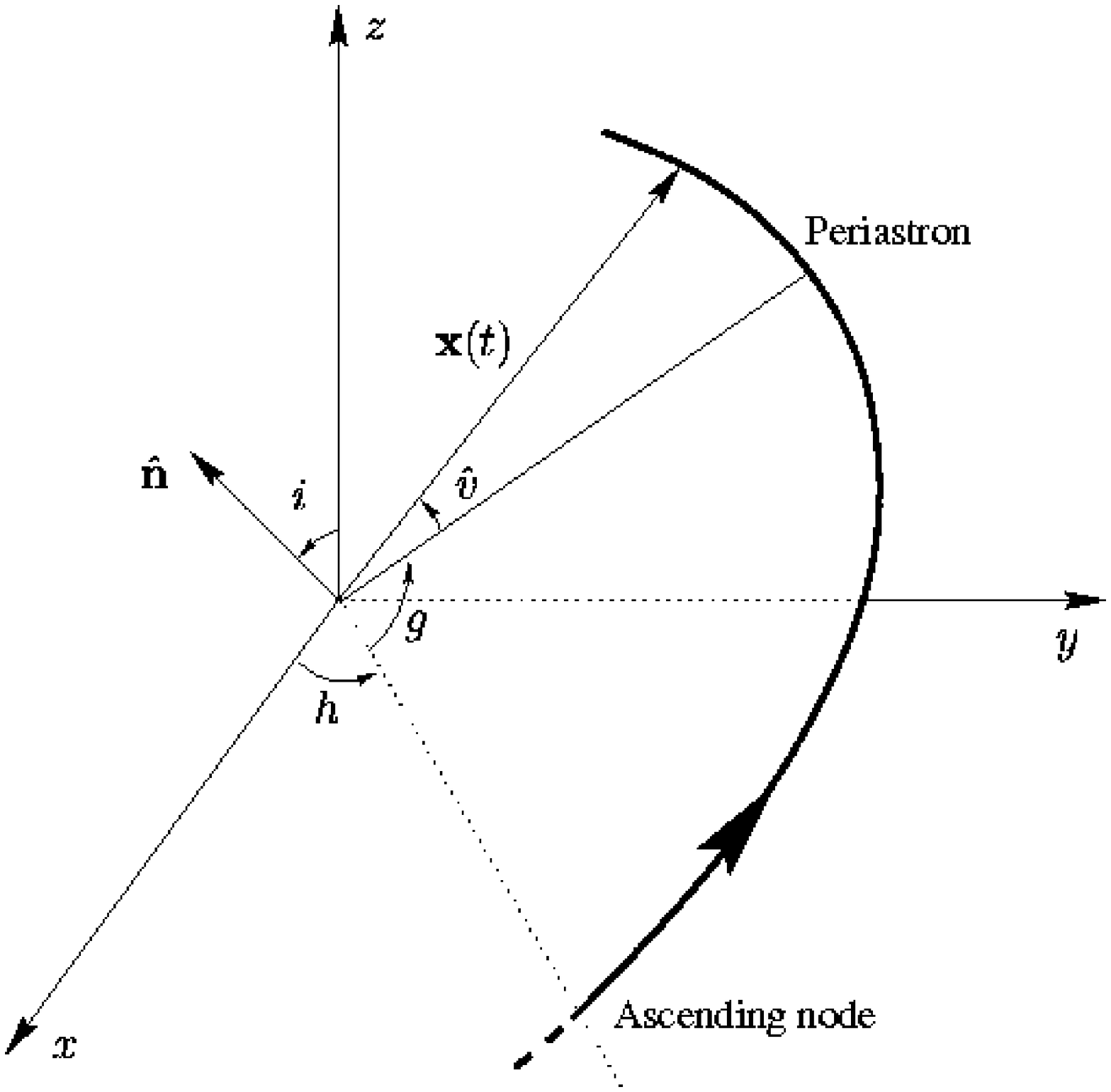}}\\
\caption[]{}
\end{figure}

\begin{figure}[h]
\epsfxsize=300pt
\makebox{\epsffile{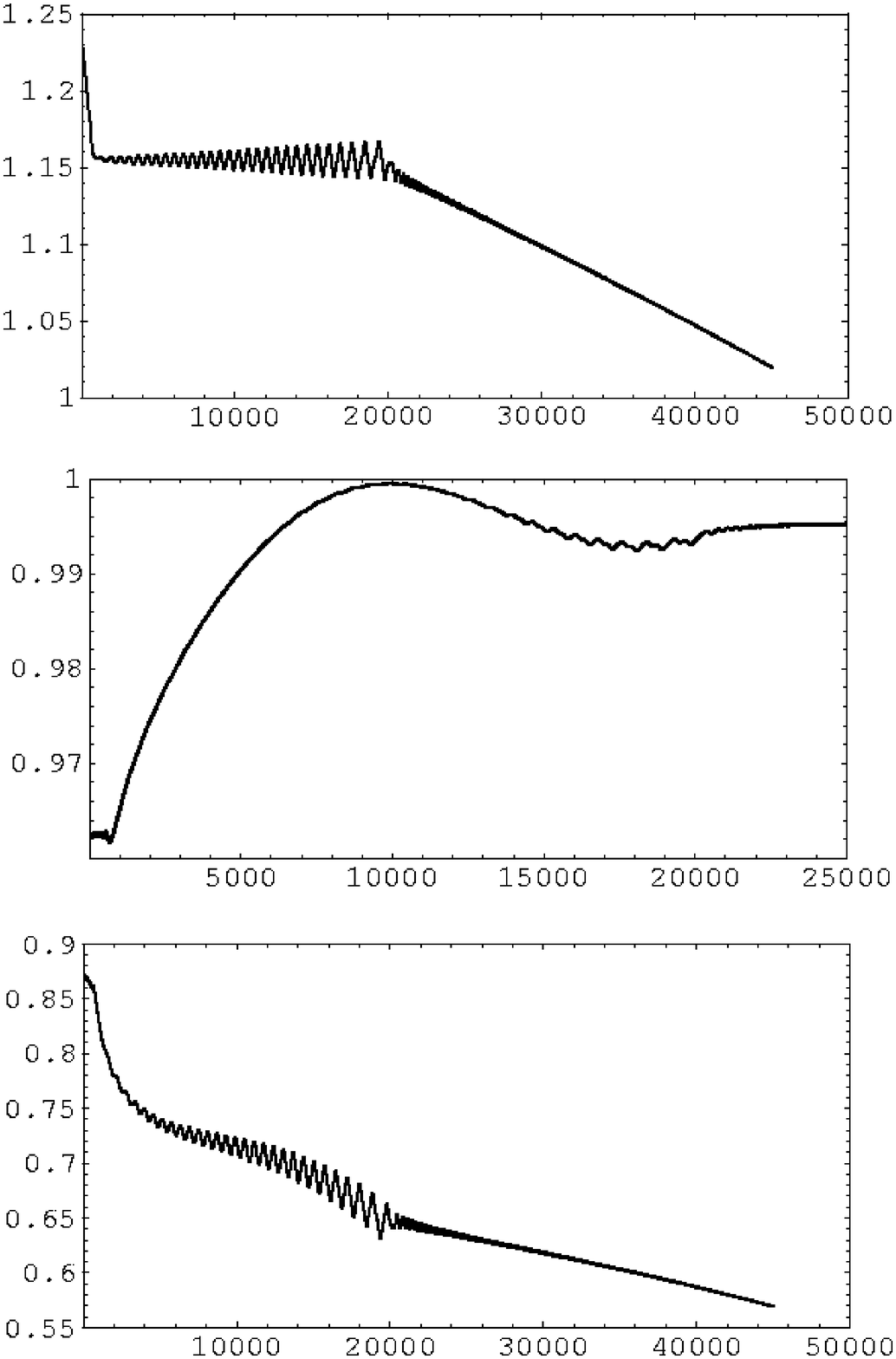}}\\
\caption[]{}
\end{figure}

\begin{figure}[h]
\epsfxsize=300pt
\makebox{\epsffile{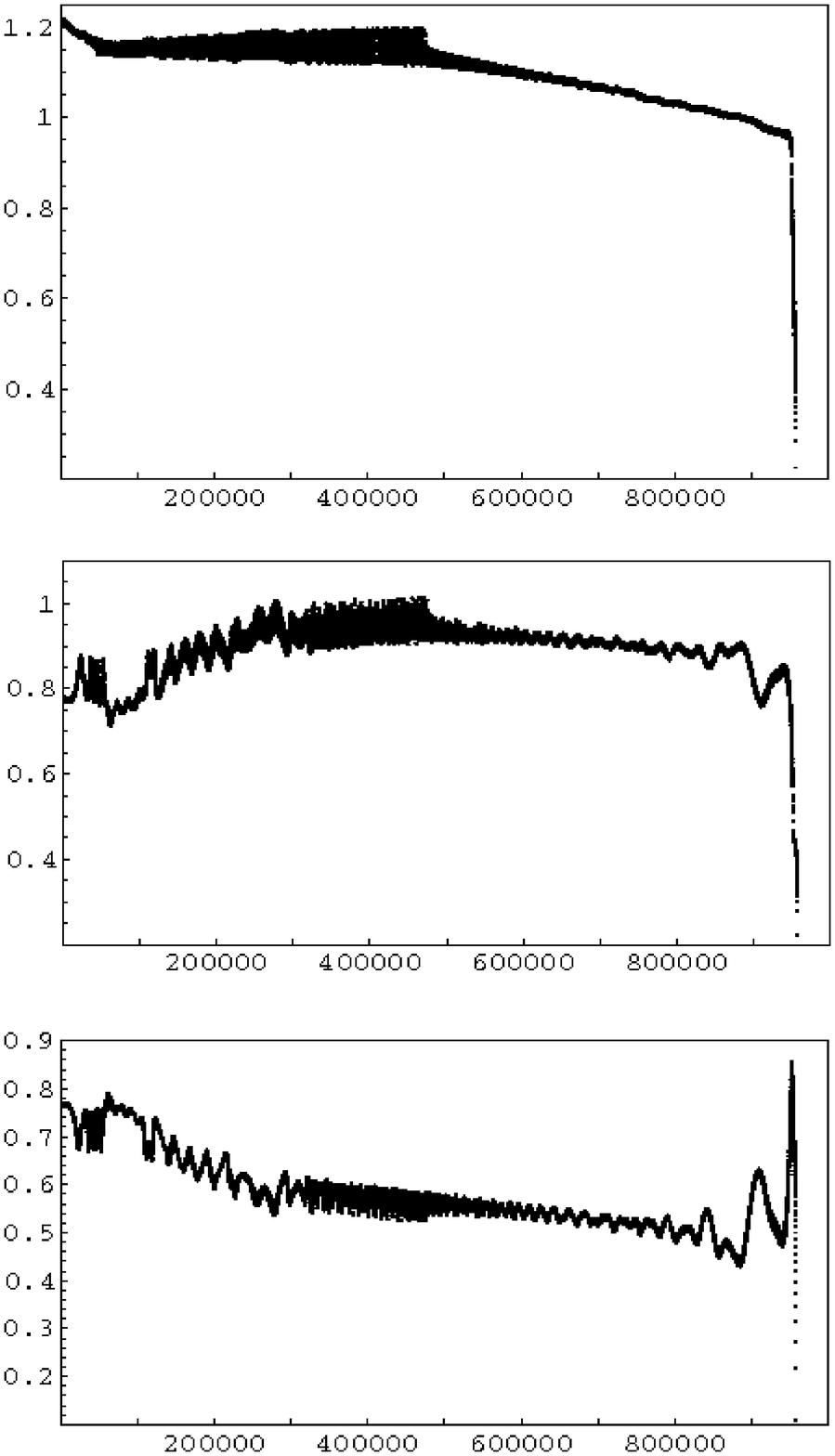}}\\
\caption[]{}
\end{figure}

\begin{figure}[h]
\epsfxsize=300pt
\makebox{\epsffile{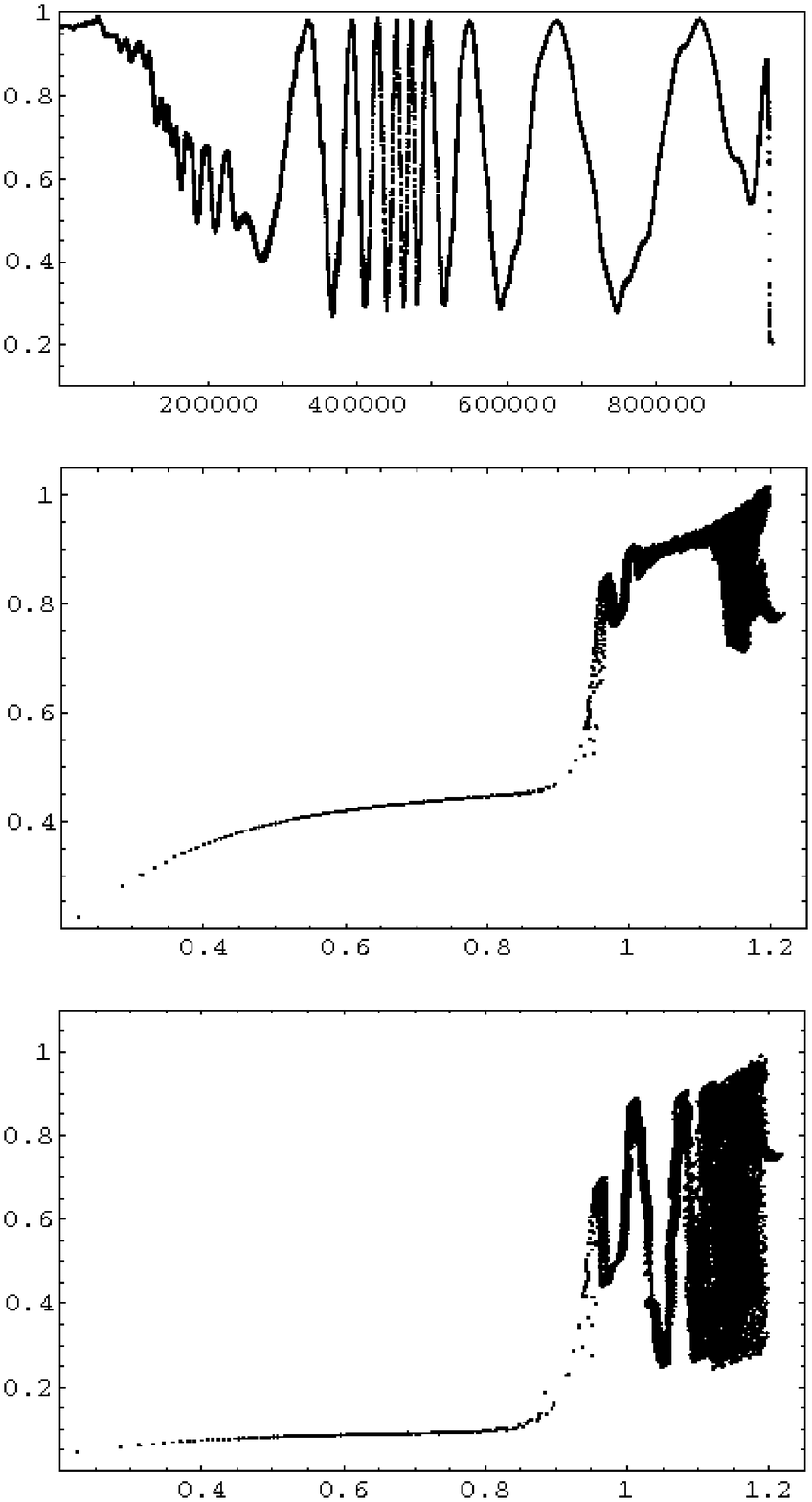}}\\
\caption[]{}
\end{figure}

\begin{figure}[h]
\epsfxsize=300pt
\makebox{\epsffile{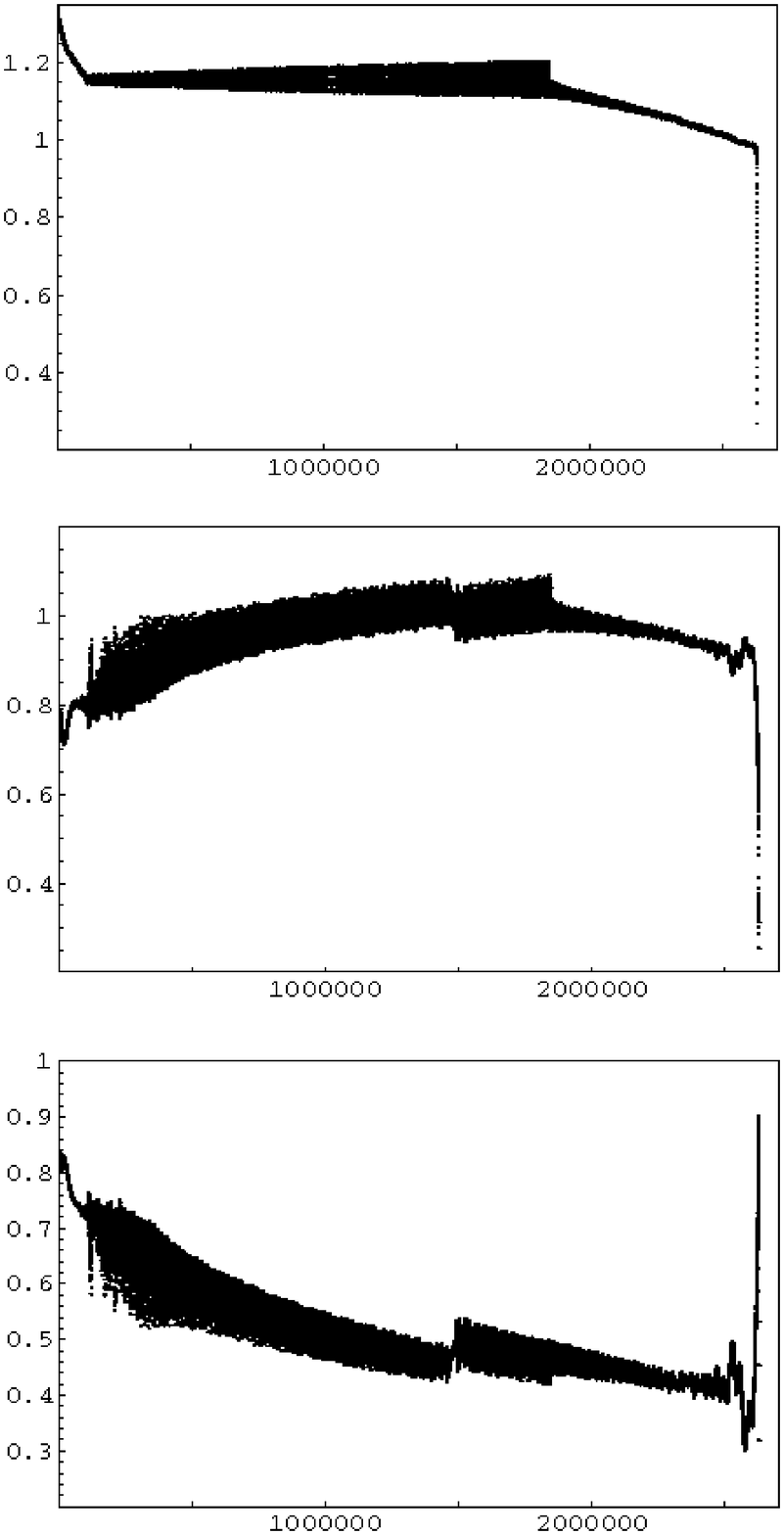}}\\
\caption[]{}
\end{figure}

\begin{figure}[h]
\epsfxsize=300pt
\makebox{\epsffile{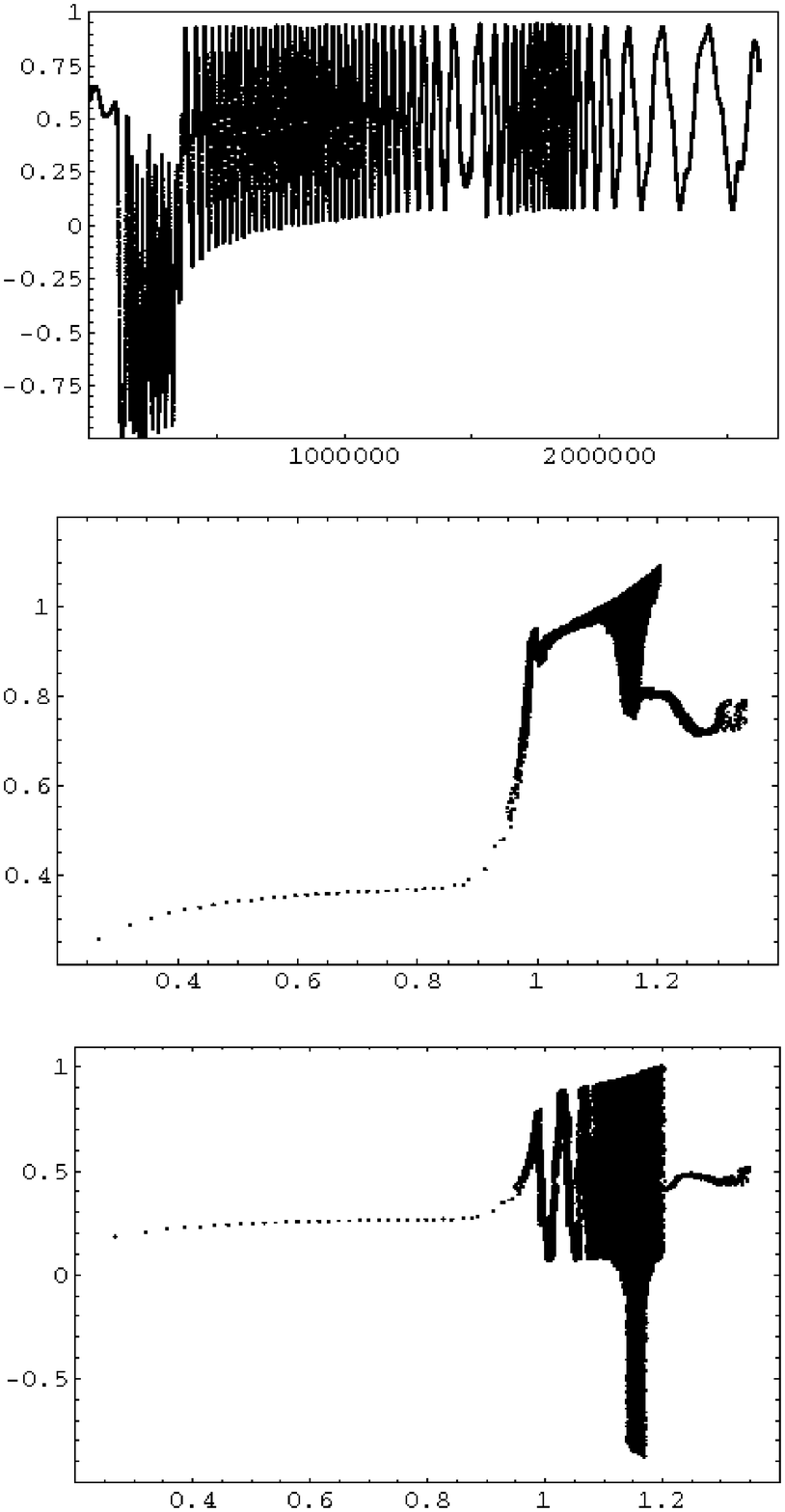}}\\
\caption[]{}
\end{figure}

\end{document}